# Dynamical Decoupling in the Presence of Realistic Pulse Errors

A. M. Tyryshkin,[1] Zhi-Hui Wang,[2] Wenxian Zhang,[3] E. E. Haller,[4] J. W. Ager,[4] V. V. Dobrovitski,[2] and S. A. Lyon[1]

[1] *Princeton University, Princeton, NJ, 08544, USA*
[2] *Ames Laboratory, Iowa State University, Ames, IA, 50011*
[3] *Fudan University, Shanghai, 200433, China*
[4] *Lawrence Berkeley National Laboratory, Berkeley, CA, 94720, USA*

**One of the most significant hurdles to be overcome on the path to practical quantum information processors is dealing with quantum errors. Dynamical decoupling is a particularly promising approach that complements conventional quantum error correction by eliminating some correlated errors without the overhead of additional qubits. In practice, the control pulses used for decoupling are imperfect and thus introduce errors which can accumulate after many pulses. These instrumental errors can destroy the quantum state. Here we examine several dynamical decoupling sequences, and their concatenated variants, using electron spin resonance of donor electron spins in a $^{28}$Si crystal. All of the sequences cancel phase noise arising from slowly fluctuating magnetic fields in our spectrometer, but only those sequences based upon alternating π-rotations about the X- and Y-axes in the rotating frame (XYXY sequences) demonstrate the ability to store an arbitrary quantum state. By comparing the experimental results with a detailed theoretical analysis we demonstrate that the superior performance of XYXY sequences arises from the fact that they are self-correcting for the dominant instrumental pulse errors in magnetic resonance experiments. We further find that concatenated sequences perform better than the periodic variants, maintaining near 100% fidelities for spin states even after several hundred control pulses. Intuitively, one would expect the instrumental error to increase with the number of pulses in the sequence but we show that the dominant first-order error does not increase when concatenating the XYXY sequence.**

Dynamical decoupling (DD) builds on the two-pulse Hahn echo technique[1] by including multiple refocusing pulses.[2] Many DD sequences have been proposed aiming at decoupling different forms of fluctuating noise.[3-8] Several experimental implementations have been reported in the context of quantum computation.[9-15] As with any open-loop control approach, DD is particularly susceptible to errors in the control pulses. Analyses of dynamical decoupling have often ignored nonidealities of any kind, only recently beginning to recognize the importance of pulse errors.[16,17] However, real systems will have pulses with both random and systematic errors. Identifying the errors specific to each experimental situation and designing DD sequences that are immune to those particular errors becomes imperative. This paper deals with ensemble measurements of electron spins where we must consider three varieties of pulse error: off-resonance errors (inhomogeneous broadening); errors in the microwave phase (rotation axis), including those arising from the leading and trailing edges of the pulses; and errors in the amplitude (rotation angle) of the pulses.

Many similar issues have been addressed in conventional magnetic resonance experiments, primarily in nuclear magnetic resonance (NMR). However, there the emphasis is usually on preserving a particular coherence, rather than all three components ($S_X$, $S_Y$, and $S_Z$) of a spin state. The canonical example is a CPMG pulse sequence,[3,4] with periodic π-rotations about the Y-axis, which preserves one ($S_Y$) spin component and sacrifices the other two components. Dynamical decoupling of a qubit does not have the luxury of preserving one component at the expense of the others.

We have studied periodic and concatenated sequences for decoupling electron spins bound to phosphorus donors in silicon. We examine two universal DD sequences, referred as XYXY and XZXZ, introduced originally in NMR by Gullion et al.,[18] and more recently in connection with DD by Viola et al.[6] Both sequences utilize π-rotations about two orthogonal axes; in the limit of ideal pulses the sequences are equivalent.[6]

In our experiments the DD pulse sequences were used to decouple magnetic field noise.[19] This field noise appears to be similar to that encountered by Biercuk, et al.[10] in their trapped ion experiments and causes a random phase fluctuation in the measured spin echo signal decays (Figure 1). As seen in Figure 1b and c, dynamical decoupling suppresses this phase noise (the results are shown for XYXY sequences but similar noise suppression was observed using XZXZ).

Figures 2a and 2b illustrate on Bloch spheres how the two DD sequences perform in the presence of pulse errors typical in spin resonance experiments. The errors in individual X, Y, and Z pulses lead the spin to evolve along a non-ideal trajectory during the pulse sequence. Nevertheless, by the end of the four-pulse cycle the XYXY sequence compensates most of the pulse errors and thus recovers the initial spin state, while the XZXZ sequence leaves a larger error.

The experimental consequences of the pulse errors are summarized in Figure 2(c-f) where fidelities are shown for different initial spin states ($S_X$, $S_Y$, $S_Z$) after applying periodic and concatenated XYXY and XZXZ pulse sequences (the Z pulses in these experiments were implemented by an X pulse followed immediately by a Y pulse). It is clearly seen that a periodic XZXZ (Figure 2c) does not equally protect all components of the magnetization. Pulse errors cause the $S_X$ state to decay rapidly, becoming nearly zero after only two cycles of the sequence (12 pulses). On the other hand, a periodic XYXY sequence (Figure 2d) treats the initial $S_X$ and $S_Y$ states essentially equally, persisting for more than 32 cycles (128 pulses).

Concatenated sequences were constructed as prescribed by Khodjasteh and Lidar.[7] Many more pulses are utilized for these sequences; 340 pulses at a concatenation level of 4, for example. As shown in Figure 2e, for XZXZ-based sequences the initial $S_X$ and $S_Y$ states are now treated on a more equal footing, however the fidelity is still significantly below unity. In contrast, the concatenated XYXY-based sequence preserves all three spin states with essentially 100% fidelity out to a concatenation level of 4 (Figure 2f).

We have performed numerical simulations taking into account the pulse errors specific to our experiments. An isolated spin model is appropriate because of the low density of magnetic $^{29}$Si nuclei and the low doping density in our $^{28}$Si-enriched sample; the residual effects of dipolar interactions will be discussed below. The spin Hamiltonian in the rotating frame is $H = \Delta\omega \cdot S_Z + \omega_1(t) \cdot (\vec{n} \cdot \vec{S})$, expressed in angular frequency units. The first term is the Zeeman interaction in magnetic field $B_0$ applied along the laboratory Z axis where only the offset, $\Delta\omega = (\gamma B_0 - \omega_0)$, appears in the rotating frame, with $\gamma$ the gyromagnetic ratio, and $\omega_0$ the resonant microwave frequency. The second term in the Hamiltonian describes resonant excitation by microwave pulses, where $\omega_1(t) = \gamma \cdot B_1$ during the microwave pulse, and $\omega_1(t) = 0$ between the rectangular pulses, with $B_1$ and $\vec{n}$ the amplitude and direction of the microwave magnetic field. For example, ideal X pulses will rotate spins by an angle $\omega_1 t_p = \pi$ (where $t_p$ is the pulse duration) around the X axis in the rotating frame, with $\vec{n} = (n_X, n_Y, n_Z) = (1,0,0)$.

We have previously characterized the pulse errors in our ensemble ESR experiments,[20] and more details can be found in the Methods section. Four types of errors can be important: (1) variations in $\Delta\omega$ for different donors in the sample; (2) rotation angle errors arising from the inhomogeneity of the microwave magnetic field, $B_1$; (3) errors in the relative phases of the nominally X and Y pulses; (4) additional phase errors from the rising and falling edges of the pulses. Careful adjustment of the spectrometer nearly eliminates (3), leaving 1, 2, and 4, as the dominant errors.

The numerical results (dashed lines in Figure 2c-f) are in excellent agreement with the experiment, using just the magnitudes of two error distributions as fitting parameters (the same across all calculations). The other error distributions were obtained independently (see Methods). In the simulations the periodic XZXZ sequence preserves only one spin state ($S_Y$) and destroys the other two states ($S_X$ and $S_Z$) after only two cycles of the sequence (Figure 2c), while the periodic XYXY sequence maintains the fidelity of all three states at a 90% level for two cycles (Figure 2d). For both sequences the concatenated versions behave better than their periodic counterparts, providing substantially improved fidelities for all three spin states (Figures 2e and 2f). The concatenated XYXY maintains nearly 100% fidelity for all spin components to a concatenation level of 4, or 340 pulses.

Further insight into the reasons why the XYXY-base sequences outperform those based on XZXZ can be gained from an analytical analysis. In the isolated spin model evolution is exclusively governed by the externally applied (control) fields. Within this approximation, an arbitrarily complex pulse sequence can be equivalently described by a single unitary operator, $U = \exp(-i\varphi(\vec{S} \cdot \vec{a}))$, which is a simple rotation by an angle $\varphi$ around axis $\vec{a}$. For periodic sequences in the presence of small pulse errors, we find that $\varphi_{XY} = 2\pi + 4(n_Y + m_X)$ for XYXY, and $\varphi_{XZ} = 2\pi + 4n_Z(1 - \cos(\Delta\omega\tau)) - 2\varepsilon_Y + 2\varepsilon_X \sin(\Delta\omega\tau)$ for XZXZ, keeping only terms to the first order in small parameters. Here, $\varepsilon_X$ ($\varepsilon_Y$) are the rotation angle errors for X (Y) pulses, $n_i$ ($m_i$) are the rotation axis errors for X (Y) pulses (see Methods for a more detailed explanation of the errors), and $\Delta\omega$ is the offset error.

The advantage of the XYXY sequences is immediately apparent from this analysis. In the absence of errors both pulse sequences produce a $2\pi$ rotation, although around different axes, $\vec{a}$ (the Y axis in case of XZXZ, and the Z axis in case of XYXY). The rotation is not exactly $2\pi$ in the presence of pulse errors. To first order, only $(n_Y + m_X)$, which is the relative phase error between X and Y pulses, contributes to the rotation error, $\delta\varphi_{XY}$. On the other hand, several different errors contribute to $\delta\varphi_{XZ}$. The microwave phase error $(n_Y + m_X)$ can be made very small through standard calibration techniques,[20,21] leaving only second order errors in $\delta\varphi_{XY}$. In contrast, with several different experimental errors contributing to $\delta\varphi_{XZ}$, it is not possible to suppress them all to the same degree in an ESR experiment. The expression for $\delta\varphi_{XZ}$ assumes that the Z-pulses are formed by X and Y pulses; the result assuming direct Z pulses is similar, with first order contributions from errors in the rotation angle and axis, and from the offset error.

The rotation error accumulates in a repeated DD sequence and primarily affects the two spin components perpendicular to the rotation axis, $\vec{a}$. In ensemble measurements, or when globally decoupling multiple qubits, different spins accumulate a different $\delta\varphi$ because of the inhomogeneities in the microwave field and resonant frequencies. After many cycles the spins spread out uniformly in the plane perpendicular to $\vec{a}$, and the ensemble magnetization decays to zero. This is just what we see in the experiment (Figures 2c and 2d), where the fidelities of the two perpendicular components ($S_X$ and $S_Z$ for XZXZ, and $S_X$ and $S_Y$ for XYXY) decay most rapidly. This decay is more rapid for XZXZ than XYXY because $\delta\varphi_{XZ}$ is larger than $\delta\varphi_{XY}$, as discussed above. On the other hand, the spin components parallel to $\vec{a}$ ($S_Y$ for XZXZ, and $S_Z$ for XYXY) are less affected, and therefore survive for many cycles.

Similar expressions can be derived for concatenated sequences. We find that $\delta\varphi$ does not increase with an increase in a concatenation level. For concatenated XYXY sequences the rotation error is always $\delta\varphi_{XY} = 4(n_Y + m_X)$ to first order, independent of concatenation level. Similarly, for concatenated XZXZ sequences we find $\delta\varphi_{XZ} = -2\varepsilon_Y$ for all concatenation levels equal to or greater than 2. The fact that the error does not increase with concatenation level, while the number of pulses increases exponentially, is quite remarkable and explains the superior performance of concatenated sequences observed in the experiment.

A few precautions are in the order. In some concatenated sequences there might be two or more Y pulses in a row (with no intervening delays). It is tempting to simply cancel pairs of adjacent identical pulses, since together they would seem to be just a 2π rotation. The solid black squares in Figure 2f are the experimental result for an initial $S_Z$-state decoupled with the XYXY sequence where such adjacent identical pulses have been cancelled. The fidelity is reduced to 77% for concatenation level 4, as compared to nearly 98% when all pulses were included. The delicate balance which allows the XYXY-based sequences to correct for the experimental pulse errors is upset by cancelling some of the pulses.

While DD sequences are effective at preserving a quantum state, they can lead to unexpected consequences. In Figure 3 we show the apparent $T_2$ of the donor electron spins as measured at various concatenation levels of the XYXY and XZXZ sequences and after repetitions of the basic sequences. The apparent $T_2$ is increasing with the concatenation level, suggesting that the effect of environmental noise is being further reduced. However, the real situation is more complex. Suspiciously, the apparent $T_2$'s obtained with DD are longer than the true $T_2$ for this sample (red line in Figure 3), known from a magnitude-detected Hahn echo experiment.[19] The echo decay is controlled by instantaneous diffusion, and such a dipole-dipole process would not be refocused by a series of ideal π-pulses. Other sequences, such as CPMG and those based on XZXZ (crosses in Figure 3) show even longer decays. Pulse errors cause unexpectedly long apparent $T_2$'s in two ways. First, non-ideal π-pulses cause the spins to develop a significant Z-component during the DD sequence (as seen in Figures 2a and 2b), even though the Z-component returns to nearly zero at the end. Since the spins have a significant average Z-projection through the course of the sequence, their apparent $T_2$ can increase, approaching $T_1$. This interpretation is confirmed by the observation that the apparent $T_2$ becomes temperature dependent, following $T_1$, while the true $T_2$ shows no such dependence over the same temperature range. Second, multiple non-ideal π-rotations lead to a partial refocusing of the dipole-dipole interactions,[22] thus lifting the limit placed by instantaneous diffusion and producing longer apparent $T_2$'s. These results imply that caution must be used in interpreting an echo decay obtained with dynamical decoupling as a transverse relaxation time. On the other hand, these sequences lead to the desirable result that a quantum state is preserved for longer periods.

In conclusion, we have shown that errors in the control pulses can significantly affect the ability of dynamical decoupling to preserve the state of a qubit. However, armed with a knowledge of the dominant pulse errors in our ensemble ESR experiments we demonstrate a pulse sequence which is capable of protecting a qubit. Comparing decoupling with the XZXZ and XYXY sequences (equivalent for ideal pulses) we find that the former rapidly destroys the qubits, while the latter preserves them for long times. These results are accurately simulated by detailed numerical calculations, and can be readily understood through a simplified analytical analysis. The XYXY-based sequences are particularly susceptible to noise in the microwave phase,[20] but that type of noise can be reduced to negligible levels in our ESR experiments. On the other hand, the XYXY sequences are resistant to our dominant errors (rotation angle and off-resonance). Other experimental situations will be dominated by other types of noise and our analytical treatment provides a way to analyze how different sequences are affected and thus to optimize them. Even the optimal concatenated sequences have undesirable side-effects since the qubits are accurately returned to their initial states only at the conclusion of the entire sequence. A qubit lying initially in the X-Y plane acquires a non-zero Z component during part of the sequence. One penalty this introduces for decoupling qubits is that it will complicate the problem of interspersing quantum operations with control pulses. Furthermore, while DD reduces the apparent decoherence it does so at the cost of decreasing the asymmetry between longitudinal and transverse relaxation. The degree to which this asymmetry can be exploited in an overlying quantum error correction scheme is thus reduced.[23]

## Methods

Experiments were performed with a Bruker Elexsys 580 spectrometer using specially modified software to allow for generating large numbers (over a thousand in some experiments) of microwave pulses, and a 20 watt continuous wave solid state microwave power amplifier (Amplifier Research) which maintained phase stability over the long pulse sequences. The spectrometer operates at X-band microwave frequency (~9.8 GHz) and a resonance magnetic field of $B_0$ ~ 0.35 T. The typical duration of a π-pulse in these experiments was 180 ns. Isotopically-enriched $^{28}$Si (~ 800 ppm residual $^{29}$Si) was used with phosphorus donor concentration of $5 \cdot 10^{14}/cm^3$ that was reduced from an initial value of $\sim 1 \cdot 10^{15}/cm^3$ by five passes of zone refining followed by floating-zone crystallization.[24] A large Si crystal (8×3×0.5 mm$^3$) was used with enough donors to enable the acquisition of an echo without signal averaging. Thus the field noise seen in Figure 1, which affects all donors equally, could be eliminated with the brute-force approach of squaring and adding the in-phase and quadrature components of the echo signal (magnitude defection).[19] We used this approach for the accurate extraction of the true $T_2$.

The periodic XYXY sequence consists of four periods of free spin evolution (each of time τ) alternating with four π-pulses, [τ – X – τ – Y – τ – X – τ – Y].[6] The Z-pulses in the XZXZ sequences were formed by an X-pulse closely followed by a Y-pulse. Thus each XZXZ cycle consists of six microwave pulses and four delays, [τ – X – τ – (XY) – τ – X – τ – (XY)].[6] The delay, τ, between pulses was 11 μs, short compared to $T_2$ and the magnetic field fluctuations. A fidelity of 1 was taken to be the intensity of the second refocused echo in a CPMG experiment measured using the same delay; the second echo in CPMG is known to correct for small pulse errors and therefore recovers the full echo intensity.[4]

Four sources of instrumental errors were considered in our simulations. The first error arises from the inhomogeneity of the local magnetic field as seen by different donors in the sample. This error translates into a non-zero resonance offset frequency, $\Delta\omega$, with a Gaussian distribution of width 50 mG (or 140 kHz) as determined from the donor ESR linewidth. The offset error causes the rotation of the spin about an axis which is tilted from the intended X or Y axis towards the Z axis in the rotating frame, and rotation through an angle, $\sqrt{\Delta\omega^2 + \omega_1^2} \cdot t_p$, which is greater than that intended ($\omega_1 t_p = \pi$).

The second error arises from the inhomogeneity of $B_1$ over the sample volume in the pulsed ESR resonator (a Bruker dielectrically-loaded cylindrical cavity, 4118X-MD5). As an approximation we assume that $B_1$ changes only along the long axis (*x*) of our sample and follows a quadratic dependence given by $B_1(x) = \bar{B}_1 + \delta B_1(1 - 3x^2/d^2)$, where $\bar{B}_1$ is the average of $B_1$ over the sample, $\delta B_1$ is the magnitude of the inhomogeneity, *x* is the distance from center of the sample, and 2·*d* is the sample length. We further assume that $B_1$ is adjusted to produce a perfect π rotation in the spin ensemble ($\gamma \bar{B}_1 t_p = \pi$). Therefore the error in rotation angle arises only from the $\delta B_1$ term, resulting in a rotation angle of $\pi + \varepsilon(x)$, where $\varepsilon(x) = \gamma \, \delta B_1 t_p (1 - 3x^2/d^2)$. With these assumptions, the distribution of rotation angle errors is described by $P(\varepsilon) = (1/2\varepsilon_0)/\sqrt{3(1-\varepsilon/\varepsilon_0)}$, where $-2\varepsilon_0 \leq \varepsilon \leq \varepsilon_0$, and $\varepsilon_0 = \gamma \, \delta B_1 t_p$ is the width of the distribution.

The third pulse error arises from imperfect relative phases of the nominally X and Y pulses. This error causes the rotation axis of the X and Y pulses to be slightly off from the intended axes, described in the rotating frame by $\vec{n} = (1, n_Y, 0)$ for X pulses and $\vec{m} = (m_X, 1, 0)$ for Y pulses, with $n_Y$ and $m_X$ being the rotation axis errors. Only the error in the relative phases of the X and Y pulses, $(n_Y + m_X)$, is important, and furthermore this error is identical for all spins in the sample. This relative phase error can be reduced to a sub-degree level through standard phase calibration techniques developed in NMR and ESR,[20,21] and therefore it is negligible in our experiments.

The fourth pulse error arises from imperfections in the rectangular shapes of the pulses, notably from the transients at the leading and trailing edges of the pulses, where the microwave amplitude and phase are not well controlled. These transients are short and constitute a small fraction (about 10%) of the total duration of the pulses, depending on the Q of the resonator. During these transients the spin is rotated about a complex, time-dependent axis (thus $\varepsilon$, $n$, and $m$ errors), and their effect depends on $\Delta\omega$, giving a distribution of the errors. We did not measure these transient errors in our experiments. Instead, to account for their effect we introduced additional rotation axis errors (such that $\vec{n} = (1, n_Y, n_Z)$ and $\vec{m} = (m_X, 1, m_Z)$ for X and Y pulses, respectively) and their distribution. Using the same arguments as above, a distribution function for $n_i$ is of the form $P(n_i) = (1/2n_0)/\sqrt{3(1 - n_i/n_0)}$, with $n_0$ defining the magnitude of the error. We assumed the transient errors to have identical magnitudes and distributions for both X and Y pulses, giving us the same form for $P(m_i)$ with $m_0 = n_0$.

In the simulations we directly solved the equations of motion for each spin, following the experimental pulse sequence, and taking into account the pulse errors discussed above. We then averaged over all spins in the sample, using the error distribution functions, $P(\varepsilon)$ and $P(n_i)$, and a Gaussian distribution for the resonant offset frequency, $\Delta\omega$. When calculating the fidelities, the simulated signal intensities were normalized by the intensity of the second refocused echo in CPMG, simulated using the same error distributions, as was done in the experiments. Our model has two adjustable parameters, $\varepsilon_0$ and $n_0$, which were determined by comparing the simulation results with the experiment. We find the best fits when using $\varepsilon_0 = 0.3$ (7.5°) for both X and Y pulses, and $n_0 = 0.12$ (3.5°) for $n_Z$ components while $n_0 = 0$ for $n_X$ and $n_Y$ components of errors in both pulses; these values were used in all numerical results presented in this paper. The estimated errors are in good agreement with those reported for similar experimental setups in other work.[20]

## Acknowledgements

We thank L. Viola, and D. Lidar for encouraging us to study dynamical decoupling and engaging in many useful discussions, J. J. L. Morton and S. Shankar for numerous useful discussions, and R. Weber and P. Hofer of Bruker Biospin for support with instrumentation. Work at Princeton was supported by the NSF through the Princeton MRSEC under Grant No. DMR-0213706 and by the NAS/LPS through LBNL under MOD 713106A. Work at the Ames Laboratory (theory and simulations) was supported by the Department of Energy - Basic Energy Sciences under Contract No. DE-AC02-07CH11358. Work at the LBNL (Si single crystal synthesis and processing) was supported by the Director, Office of Science, Office of Basic Energy Sciences, Materials Sciences and Engineering Division of the US Department of Energy (DE-AC02-05CH11231).

Figures

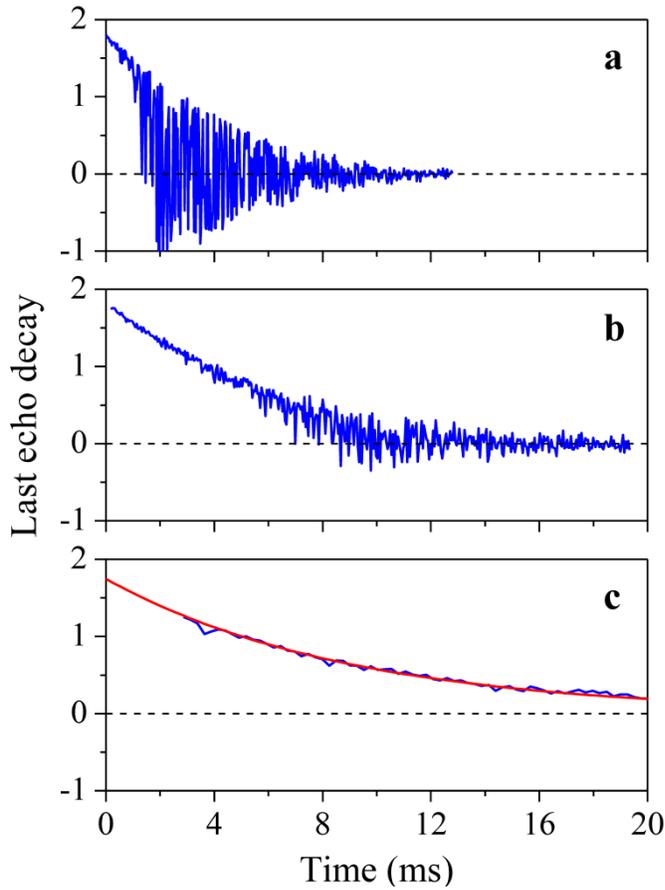

**Figure 1. Phosphorus donor spin echo signal decays in isotopically-enriched $^{28}$Si:** The amplitude of the in-phase component of the microwave spin echo signal from electrons bound to donors held at 8 K is shown for different levels of dynamical decoupling. (a) The Hahn echo decay shows substantial "phase noise" developing at times longer than 1 ms: this noise originates from the fluctuations of the magnetic field, $B_0$.[19] The power spectrum of the field noise (T$^2$/Hz) varies approximately as $1/f^2$ with an amplitude of ~50nT/√Hz at 10 Hz. (b) The onset of the phase noise shifts to longer times (> 4 ms) and the noise magnitude is reduced when using a second level concatenated XYXY sequence. To obtain these data the number of pulses in the sequence is fixed while the time delay between pulses is increased. An echo is formed at the end of the sequence, and the amplitude of the in-phase component of the echo is plotted as a function of the total time since the initial π/2 pulse. (c) The phase noise is completely suppressed with a fourth level concatenated XYXY sequence. The red line is an exponential fit to the echo decay.

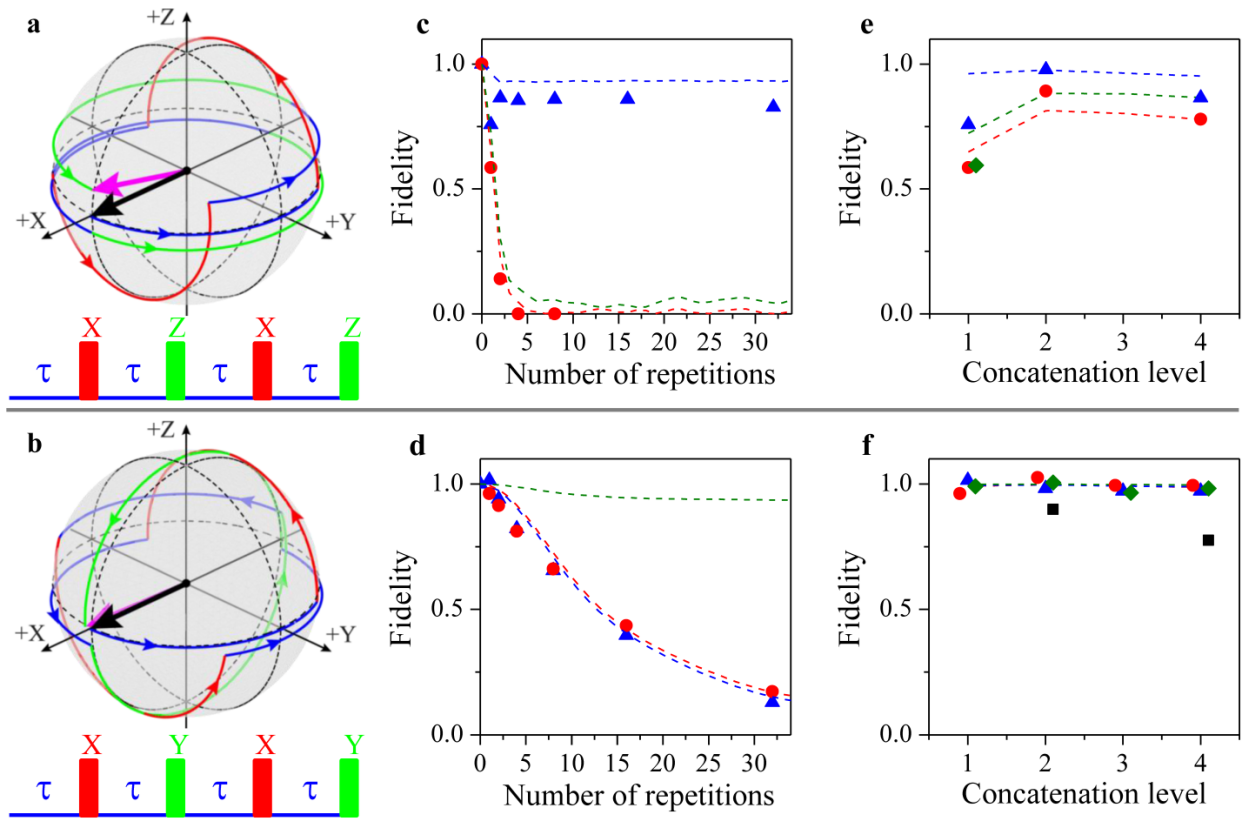

**Figure 2. Comparison of spin evolution and fidelities for XZXZ (upper panel) and XYXY (lower panel) dynamical decoupling sequences.** (a and b) Spin evolution, shown on a Bloch sphere in the rotating frame, of a single period of the XZXZ and XYXY pulse sequences, respectively, starting from an initial $S_X$ spin state (bold black arrow) and evolving into the final state (bold pink arrow). The step-by-step trajectory are traced by the colored curves: blue represents the free evolution of the spin state between microwave pulses; red curves are spin rotations by X pulses, and green curves are spin rotations by Z pulses (upper, XZXZ panel) or Y pulses (lower, XYXY panel). Pulse errors typical for our ESR experiments have been assumed: offset error, $\Delta\omega = 30$ kHz; rotation angle errors, $\varepsilon_i = 5$ degrees, for all three X, Y, and Z pulses; rotation axis errors, $n_Z = 3$ degrees, in X and Y pulses, and $n_X = 3$ degrees in Y and Z pulses. The free evolution period between pulses is 11 μs. Individual pulse errors cause a substantial deviation of the spin state from an ideal trajectory (free evolution would always lie on the equator) during the sequence. Nevertheless, after the complete cycle the XYXY (b) sequence compensates nearly all the errors and returns the spin close to its initial state. In contrast, the XZXZ (a) recovers the state with a substantial error, and this error will accumulate after repeating the XZXZ sequence. (c - f) The measured fidelities of the recovered spin states are shown for periodic XZXZ (c) and XYXY (d) sequences, plotted as a function of the number of repetitions of each sequence; and for

concatenated XZXZ (e) and XYXY (f) sequences, plotted as a function of concatenation level from 1 to 4. The fidelities were calculated as the projection (probability) of the final state on the initial state using the prescription of Ref. [25]. Three initial spin states were considered: $S_X$ (red circles) and $S_Y$ (blue triangles) were measured for all the sequences, and $S_Z$ (green diamonds) was measured for the concatenated XYXY sequences. Also shown in (f) are the measured fidelities for initial $S_Z$ states (black squares) when using the concatenated XYXY sequences but with adjacent identical pulses removed. All fidelities were measured at times short compared to $T_2$ to isolate the contribution of pulse errors to the echo decays. Dashed curves represent simulated fidelities using the numerical model described in the text and taking into account the individual pulse errors and their distributions within a spin ensemble, with $\varepsilon_0 = 7.5$ degrees and $n_0 = 3.5$ degrees (the $n_z$ component), and $\Delta\omega = 140$ kHz.

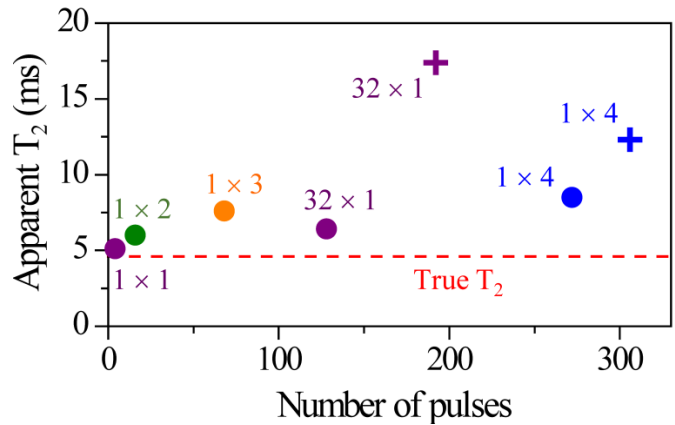

**Figure 3. Apparent $T_2$'s using XYXY and XZXZ sequences.** Measured $T_2$ times are shown for various concatenation levels of the XYXY (circles) and XZXZ (crosses) sequences, and also after repetitions of the basic sequences. The data are plotted as a function of the total number of pulses in the sequence (remembering that the basic XZXZ has 6 pulses). The label "n×m" indicates "n" repetitions of the "m$^{th}$" concatenation level. For these experiments all the delays (τ) between pulses were varied together to obtain the last echo decay as a function of the total time. Magnitude detection was used to eliminate the effect of magnetic field noise and allow unambiguous determination of the echo intensity since the shorter sequences only partially decouple the noise. The horizontal dashed (red) line indicates $T_2$ (4.6 ms) measured in a Hahn echo experiment using magnitude detection. We take it to be the true $T_2$ without pulse errors and it is known to be limited by dipole-dipole interactions between donor spins (instantaneous diffusion).[26]